# The Challenge of Value Alignment: from Fairer Algorithms to AI Safety

Iason Gabriel and Vafa Ghazavi

1. <u>Introduction</u>

There has long been a view among observers of artificial intelligence (AI) research, often expressed in science fiction, that it poses a distinctive moral challenge. This idea has been articulated in a number of ways, ranging from the notion that AI might take a 'treacherous turn' and act in ways that are opposed to the interests of its human operators, to deeper questions about the impact of AI on our understanding of human identity, emotions and relationships. Yet over the past decade, with the growth of more powerful and socially embedded AI technologies, discussion of these questions has become increasingly mainstream. Among topics that have commanded the most serious attention, the challenge of 'value alignment' stands out. It centres upon the question of how to ensure that AI systems are properly aligned with human values and how to guarantee that AI technology remains properly amenable to human control. In 1960, the mathematician and founder of cybernetics, Norbert Wiener, articulated a version of this challenge when he wrote that, 'if we use, to achieve our purposes, a mechanical agency with whose operation we cannot interfere effectively… we had better be quite sure that the purpose put into the machine is the purpose which we really desire' (1960). More recently, the prominent AI researcher Stuart Russell has warned that we suffer from a failure of value alignment when we 'perhaps inadvertently, imbue machines with objectives that are imperfectly aligned with our own' (2019, 137).

This chapter explores the questions posed by AI and alignment with human values in more detail. It begins by looking at foundational questions about the relationship between technology and value. Far from existing in a vacuum, we note that there has long been interest in, and concern about, the potential of technology to 'lock-in' different values and kinds of authority relationship. Moreover, recognition of this fact has often been accompanied by an understanding that it places a degree of responsibility on designers, and recognition that we need specific methodologies to help align technology with visions of the social good that receive widespread endorsement. With this framework in place, the second part of the chapter looks at the question of AI technology in more detail. In particular, we ask: is something special about AI that makes questions about value more complicated or acute? Here, we answer in the affirmative: while there are clear continuities with technologies that have come before, the combination of intelligence and autonomy demonstrated by modern AI systems gives rise to new challenges. This is true both from a normative perspective, given that we are able to encode a richer set of values in AI systems than in more simple artefacts, and also from a technological

perspective – where greater scope of action and intelligence create new challenges from the perspective of alignment and control. The third part of this chapter looks at work being undertaken by technical AI researchers to address the challenge of alignment over the long run and at discussions taking place within that community. Finally, with this understanding in place, we return to earlier observations about the relationship between technology and value, and ask how this philosophical and sociological work might contribute to our understanding of AI alignment. In this context, we note that while most technical discussion of alignment focuses on one-person-one-agent scenarios, we also need to think seriously about social value alignment. This would involve aligning AI systems with a range of different voices and perspectives.

## 2. Technology and Values

Although the challenge of value alignment has arisen first and foremost in discussion of AI systems, the notion that technology has a moral valence and moral consequences has a long intellectual lineage. From a philosophical vantage point, value generally refers to what ought to be promoted in the world, encompassing concepts such as autonomy, justice, care, well-being and virtue. This normative conception of value has deep philosophical roots, and can be contrasted with instrumental value used to price goods in markets (Satz 2010). Outside philosophy, there is also considerable intellectual inquiry into the relationship between technology and values. In the context of technological design, a working definition of 'value' has been offered by Friedman and Hendry as 'what is important to people in their lives, with a focus on ethics and morality' (2019, 24). Moreover, the entire interdisciplinary field of science and technology studies (STS) builds upon the insight that values tend to be embedded in technologies, with ramifications for technologists and society, including through the impact on norms and ways of life. As Sheila Jasanoff puts it, 'far from being independent of human desire and intention, [technologies] are subservient to social forces all the way through' (2016, 18). Beyond this, technological artefacts have the potential to 'lock-in' or manifest certain values in a variety of ways.

One famous example of this phenomenon can be found in Langdon Winner's seminal article, 'Do Artifacts Have Politics?' (1980). In it he cites the case of Robert Moses, whose twentieth-century designs for New York City contributed to deepening racial stratification by creating low-hanging bridges that limited public transport flows from poorer predominantly non-white neighbourhoods to more affluent public spaces.[1] Another example of 'design with a purpose' can be found in the city plans of Baron Haussman who redesigned the streets of Paris after the events of the French revolution, so that it contained open boulevards that facilitated troop movement and suppressed the possibility of protest (Scott, 2020). In these ways, technical artefacts may have 'political properties' embodying specific forms of power and authority. Moreover, Winner suggests that there is a temptation, with the creation of powerful tools, for the ends to be adapted to means (rather than the other way around), and for technologies to draw forth certain modes of social organization. In the context of discussions around AI, Yuval Hariri has recently made a version of this argument, suggesting that the need for large datasets and computer power favors centralized forms of political authority and governance (2018).[2]

---

[1] This was also a problem in other cities in the US. For example, the Interstate 20 in Atlanta was deliberately plotted, in the words of Mayor Bill Hartsfield, to serve as 'the boundary between the white and Negro communities.' Black neighborhoods, he hoped, would be hemmed in on one side of the new expressway, while white neighborhoods on the other side of it would be protected (Kruse, 2019)

Viewed from a philosophical vantage point, it should also be clear that technology and value are intimately connected. Crucially, technological artefacts, whether in the form of transport systems, medical innovations, communications devices, energy facility design, or the development of home computers, shape and influence the choice architecture within which individual decisions are subsequently made. This fact is easily obscured by an agent-centric view of ethics. However, it should be clear, upon reflection, that new technologies make some outcomes more likely and some outcomes less likely to occur, that they create new possibilities, and that their creation will sometimes exclude certain possibilities from being realised altogether. In this sense, technology often has a profound effect on the states of affairs that we are able access on an individual or collective basis, and on the states of affairs that we are likely to pursue. Thus, regardless of the metaethical stance we take – including our view about where value ultimately resides – no technology of scale will be 'value neutral' in the sense of leaving this balance of possible outcomes unaffected.[3]

This insight about what we term the 'technology-value nexus' has a number of important consequences. First, it suggests that technologists are themselves engaged in a world-making activity and that a level of responsibility necessarily attaches to the process of technical innovation and design. For even if Heidegger (1954/2013) is correct, and the cumulative growth of technology ultimately represents an 'enframing' or drawing forth of possibilities that are hard to fully foreshadow or understand, some measure of foresight is possible – creating the opportunity for responsible agency and direction in this domain. Second, in order to discharge this responsibility successfully, and guard against moral error, we need methods to ensure that the design of technology is congruent not only with the personal moral beliefs of designers or engineers but also with a vision of flourishing that is widely endorsed and sensitive to the needs of different communities.[4] Since the early 1990s, an approach known as 'value sensitive design', which draws on fields such as anthropology, human-computer interaction, philosophy, and software engineering, has actively sought to bring values into technological processes (Friedman and Handry 2019). Key methods for doing so include techniques such as stakeholder analysis and citizen consultation, both of which emphasize the importance of including the perspective of those who are significantly affected by technological innovation (Martin, 2020). Third, it suggests that technologists need to think seriously about these questions early on, including whether to develop certain technologies at all. It is almost always beneficial to exercise ethical foresight at this point, when there is greater latitude of choice and before various path-dependencies have set in.[5]

---

[2] For a contrasting view see Andrew Trask (2020).

[3] This claim holds true both for person-affecting and non-person-affecting theories of value (Parfit, 1997).

[4] Admittedly, this claim rests on a set of prevailing metaethical views which allow for the possibility of moral error. Certain forms of anti-foundational metaethics or error theory dispute this claim. Yet, a more interesting question still, concerns what happens when society as a whole is in a state of error. What should we say about the design of technology then? Are there ways that entire societies can uncover moral blindspots and ensure that they do not carry over into the design of AI systems? These questions take us into the domain of what Allen Buchanan terms 'social moral epistemology' (2002). In his view there are certain institutions and social practices that society may adopt that reduce the risk of systematic moral error.

### 3. Is A.I. Special?

With this understanding in place, we can now ask whether AI is 'special' when considered from the standpoint of the technology-value nexus. Is AI just like the other technological artefacts we have mentioned, or does it differ in some important way? To make progress on this question we start with a brief overview of what AI is, or has come to mean, at the present moment. Building upon this foundation, we then consider what unique properties AI systems might possess.

What is A.I.?

The term 'artificial intelligence' commonly refers both to a quality of computerized systems and to a set of techniques used to achieve this capability, most often machine learning (ML). 'Intelligence' can be understood in this context to refer to an agent's ability to adapt and 'achieve goals in a wide range of environments' (Legg and Hutter 2007, 402). This instrumental conception of intelligence represents only one view drawn from a family of definitions one might endorse (Dignum, 2019; Cave, 2020). However, within the vocabulary of computer science 'artificial intelligence' tends to refer primarily to models or agents that perceive their environment and make decisions that maximize the chance of achieving a goal. Indeed, Stuart Russell articulates this view clearly when he writes that 'machines are intelligent *to the extent* that their actions can be expected to achieve their objectives' (2009, 19).

In this context, ML refers to a family of statistical or algorithmic approaches that can be used to train a model so that it learns to perform intelligent actions – sometimes from scratch. The discipline of ML encompasses a variety of different approaches. One branch called 'supervised learning' focuses on training a model to identify and respond to patterns in labelled datasets. Supervised learning is at the heart of many real-world applications of AI including automated image recognition, disease diagnosis, the development of financial trading strategies, and the creation of job recommendation systems. 'Unsupervised learning', by way of contrast, aims to uncover patterns in un-labelled data and to perform tasks, such as the discovery of fraudulent transactions, on that basis. When run on sufficiently powerful hardware, both techniques allow models to learn from experience without using explicit instructions. In this regard, ML systems differ from earlier 'expert system' models such as the Chess-playing computer Deep Blue, which relied upon an intricate set of hand-crafted rules and instructions to defeat the reigning Chess champion Gary Kasparov in 1996. Innovations in ML, the collection of vast datasets, and the growth of computing power have together fostered many recent AI breakthroughs.

---

[5] Indeed, the discipline of value sensitive design has engaged with AI from its beginnings (Friedman and Kahn 1992).

Looking forward, one particularly promising approach for more advanced forms of AI is reinforcement learning (RL). RL agents usually contain four key elements: a policy which defines the agent's way of behaving at a given time; a reward signal which defines its goal; a value function which estimates the long-term value of different states of affairs; and a model of the environment which allows the agent to make predictions about how the environment will respond to its decisions (Sutton and Barto 2018, 6-7). RL agents then learn what to do by trying to maximise a numerical reward signal that they receive from their environment. They do this by engaging in exploration and revising their policies along the way. Sophisticated RL agents have proved particularly adept at game-playing, mastering the ancient Chinese board game of Go (which had previously been thought to be computationally intractable because of the vast number of possible moves) as well as real-time computer strategy games such as Defence Of The Ancients (DOTA) and StarCraft II.

The Potential Uniqueness of A.I. Systems

Many of these innovations are impressive in their own right. However, when considered from the standpoint of the technology-value nexus, do they represent some fundamental departure from previous systems or a significant point of differentiation? In many ways, the answer is no. Concerns over injustice, safety, and unintended consequences exist for ML algorithms as they do for other technologies. Moreover, the central challenge discussed so far, which centres on the potential for technology to lock-in or manifest a particular set of values, has clearly spilled over into the domain of ML where it is usually discussed under the guise of 'algorithmic bias'.

Recent analysis has identified numerous cases where algorithmic tools or models have come to reflect forms of bias that arise from the data they were trained on – or from the way that data was curated and labelled. In the case of natural language processing, algorithms learned to associate certain job types with gender stereotypes, leading to biased predictions that disadvantaged women. Historically compromised data has also led to racially-biased recommendations in the context of the criminal justice system, both when it comes to parole recommendations and with predictive policing (Angwin et al 2016; Lum and Isaac 2016). And there are many examples of ML systems performing worse for minorities or groups who sit at the intersection of different forms of disadvantage. In particular, automated facial analysis algorithms (Buolamwini and Gebru 2018) and healthcare diagnostics (Obermeyer, 2019) have tended to perform poorly for women and non-white sections of the population. When these decision-making systems are used in socially important contexts such as the allocation of healthcare, education, credit, they can compound disadvantage by obscuring its origin, extending its influence over time, and creating forms of automated discriminaton that are difficult to address (Eubanks 2018). Moreover, in each case, there is a clear sense that the technologies are not aligned with key values such as the requirements of fairness, or with what we as a society want them to do. Thus, they embody what we might term 'social value misalignment'. To address these shortcomings, there is now an active fairness, accountability, and transparency research community (Corbett-Davies, 2018; Benjamin, 2019; Selbst, 2019; Abebe, 2020).

At the same time, the fact that this is an emergent field of academic study testifies to the fact that these problems are often particularly acute for AI, and that moreover – there are certain features of AI systems that make the task of social value alignment distinctive and particularly challenging. Indeed, the moral questions raised by AI alignment are *not simply* those that are intrinsic to the creation of large-scale technological systems, or technologies that occupy public roles, functions, and places. Some of these unique challenges are to do with complexity and opacity of the systems involved: once an algorithmic model has been trained, it can be hard to say why it decided upon the action or recommendation it arrived at. Others challenges are more precisely tied to the notion of machine intelligence and autonomy in decision-making – to the idea that AI systems can make decisions or choices that are more meaningful than those encountered by technologies in the past.

To see this clearly, we can start by noting that simple technological artefacts, such as hammers or pencils, are not able to respond to their environment – let alone make decisions. Against this backdrop, Daniel Dennett suggests that even the existence of a simple 'switch' that can be turned on and off by some environmental change marks a *degree of freedom*, which is 'a level of interactivity that can be influenced and needs to be controlled' (2003, 162). More complicated artefacts have additional degrees of freedom in case where 'there is an ensemble of possibilities of one kind or another, and which of these possibilities is actual at any time depends on whatever function or switch controls this degree of freedom' (ibid.). As these switches or nodes proliferate, they form 'larger switching networks, the degrees of freedom multiply dizzyingly, and issues of control grow complex and nonlinear' (ibid.). For Dennett, these properties are evidenced by biological organisms of sufficient complexity, including humans. However, they also are also found in the networks used by AI researchers to create models that learn from their environment and optimise for objectives. As a consequence of this design feature, artificial agents can learn new mappings between inputs and outputs, coming up with solutions (or failure modes) that sometimes surprise their human designers.

Partly as a consequence of this freedom, it is possible to 'load' a richer set of values into AI systems than with more simple artefacts. This can be seen, for example, in the case of self-driving cars that have to navigate the world successfully, while managing complex trade-offs in emergency situations. It is also reflected by the fact that the behavior of AI systems is best understood by adopting what Dennett terms 'the intentional stance' (2009). Whereas it is possible to understand the behaviour of simple artefacts either by reference to mechanical explanations or design principles, it is most useful to think of AI systems as rational agents that have goals and intentions.[6] Moreover, whereas simple artefacts can 'be seen to derive their meaning from their functional goals in our practices, and hence not to have any intrinsic meaning independent of our meaning', AI systems have trajectories that 'can unfold without any direct dependence on us, their creators, and whose discriminations give their internal states a sort of meaning to them that may be unknown and not in our service' (ibid). Indeed, as these systems make decisions previously reserved as the domain of human control, we are led to ask questions about where responsibility for these behaviors ultimately resides and how to ensure they are subject to meaningful control.

One answer to the first question has been provided by Floridi and Sanders who argue that AI systems can also be 'moral agents' (2004). As with Dennett, the level of abstraction needed to analyse their behaviour plays a central role in the argument. Indeed, for these authors the (a) interactivity, (b) autonomy and (c) adaptability of AI systems, make it possible for them to have this status (Floridi and Sanders 2004, 357-58). In this context, interactivity indicates that an agent and its environment can act upon each other. Autonomy 'means that the agent is able to change state without direct response to interaction' meaning 'it can perform internal transitions to change its state' which gives it 'a certain degree of complexity and independence from its environment' (ibid.). Adaptability refers to the agent's potential, through interactions, to change the 'transition rules' by which it changes state. Taken together, these properties mean that an agent can learn its own mode of operation based on experience, and could, according to Floridi and Sanders, be made accountable for moral action.

However, even without this strong claim about moral responsibility, it seems clear that AI models have agential properties that manifest to a higher degree than in non-AI systems. This insight has implications for the kind of normative questions we can meaningfully ask about AI. For example, we can ask: which principle of values should we encode in AI – and who has the right to make these decisions – given that we live in a pluralistic world that is full of competing conceptions of value (Gabriel, 2020)? Can AI be made so that it is kind or compassionate? Can it demonstrate care for human beings and sentient life? In this case, the moral qualities of 'compassion' or 'care' refer not to a subset of decisions made within a defined action space, but rather to a standing disposition, to a set of counterfactual truths about what an agent would do across a variety of circumstances and contextual variations.[7] Questions of this kind, make little sense for simpler technologies. At most, we might ask whether a car, transport system, or simple computer program, was *designed* in a way that demonstrated compassion or concern for users and non-users of the technology. But with AI systems, the locus of meaningful analysis shifts to the qualities of agents themselves.

---

[6] This then leads to the heavily disputed question of whether these artefacts can *really* be said to have goals and intentions. Dennet suggests that they can. Against this view, one might argue that these artefacts do not obviously have the quality of possessing 'mind'. Both perspectives are compatible with the view we are advancing here.

[7] This idea that the quality of compassion should be treated as a fixed disposition draws heavily upon the notion of a 'virtue' used in Virtue Ethics. As Rosalind Hursthouse writes on this point, 'Given that the virtuous disposition is multi-track, no virtue ethicist would dream of making "the fundamental attribution error" and ascribing honesty or charity to someone on the basis of a single honest or charitable action or even a series of them' (2006, 102)

4. Technical Approaches to Value Alignment

If the preceding analysis is correct, then the alignment of powerful AI systems requires interdisciplinary collaboration. We need a clearer understanding both of the goal of alignment *and also* of the technical means available to us for implementing solutions. In this regard, technical research can provide us with a more precise understanding of the challenges we face with AI and about the kind of answers that are useful. Meanwhile, from the philosophical community and the public, we need further direction and guidance about the goals of alignment and about the meaning of properly aligned AI. In the spirit of mutual endeavour, this section looks at technical aspects of the alignment challenge, including methodologies for achieving AI alignment, concrete problems encountered to date, and proposals about how to ensure AI systems stay aligned even if their intelligence one day significantly exceeds our own.

Top-Down and Bottom-Up Approaches

When it comes to strategies for creating value-aligned AI, Wallach and Allen distinguish between 'top-down' and 'bottom-up' approaches (2009). Top-down approaches to alignment start by identifying an appropriate moral theory to align with and then designing algorithms that are capable of implementing it. With this approach the designer explicitly sets an objective for the machine from the outset based on some moral principle or theory which they would like to operationalize. By way of contrast, bottom-up approaches do not require the specification of a full moral framework. Instead, they focus upon the creation of environments or feedback mechanisms that enable agents to learn from human behavior and be rewarded for morally praiseworthy conduct. Each approach brings with it technical and normative challenges.

To start with, top-down approaches are based on the possibility that ethical principles or rules can be explicitly stated, that these principles can be expressed in computer code, and that following these principles constitutes ethical action (Wallach and Allen 2009, 83). The relevant ethical principles could derive from religious ideals, moral codes, culturally endorsed values, or philosophical systems (Allen et al 2005, 150). This approach to alignment has also been explored in science fiction, with Isaac Asimov's 'Three Laws of Robotics' serving as a classic illustration. The rules he proposed (i) banned robots from injuring humans; (ii) insisted they obey humans – except where this would violate (i); and (iii) stipulated a robot must protect its own existence – as long as this didn't violate (i) or (ii).

If it can be made to work, a top-down approach has certain advantages: the rules it relies upon could in principle be widely known and understood – and they could be designed to target undesirable behaviors (such as killing or stealing). However, as Asimov's stories illustrate, rules can also come into conflict with each other, producing 'computationally intractable situations unless there is some further principle or rule for resolving the conflict' (Allen et al 2005, 150). More importantly still, this approach appears to require us to identify and specify the correct moral framework for AI upfront. If this is the case, then it forces us onto the horns of a dilemma: either we must proceed on the basis of our own personal moral beliefs (which could easily be mistaken) or we need to identify public principles for AI that are, in practice, difficult to come by. So far, variants of utilitarianism have tended to be the preferred option for engineers, given the apparent compatibility of the theory with optimization-based learning (Roff, 2020; Russell, 2019). Yet this trajectory is problematic if our ultimate goal is social value alignment (i.e. for the alignment of AI systems with values that are widely endorsed).

In the light of how hard it is to identify and encode appropriate moral goals, some researchers have instead pursued a 'bottom-up' approach to alignment that seeks to infer human preferences about values from observed behavior or feedback. In fact, there is a specific branch of RL called inverse reinforcement learning (IRL) which appears well-suited to the task. IRL systems do not directly specify the reward function that the agent aims to maximize. Rather, in these models, the reward function is treated as an unknown which must be ascertained by the artificial agent. More precisely, the agent is presented with datasets (including potentially very large ones), environments, or set of examples (such as the conduct of human experts), and focuses on 'the problem of extracting a reward function given observed, optimal behaviour' (Ng and Russell 2000, 663). The goal of the exercise is then to infer or understand human preferences through observation and to align with them, rather than pursuing an independently specified goal or outcomes.

However, the bottom-up approach to value alignment also encounters challenges. In certain respects it tends to be more *opaque* than ML systems where the reward is clearly specified. With IRL even if the agent appears to be acting in a moral manner, it will be hard to know what precisely it has learned from the dataset or examples we have shown it. Moreover, important normative questions still need to be addressed (Gabriel, 2020). As the fairness, accountability and transparency research community has demonstrated, the salient question immediately becomes what data to train the system on, how to justify this choice, and how to ensure that what the model learns is free from unjustifiable bias.

One interesting data-point for bottom-up approaches comes from the 'Moral Machine' experiment which crowd-sourced the intuitions of millions of people about moral trade-offs encountered by autonomous vehicles (Awad, 2018). Ultimately, the result of the study was inconclusive despite its scale. What it revealed was a set of noisy preferences in this area, some obvious tendencies (e.g. to value many lives over fewer lives), and some ethical variation across cultures, including a propensity to accord more ethical weight to the lives of higher status individuals in poorer countries. The study therefore raises deep questions about the coherence of everyday moral beliefs and perspectives across populations, and also about the value of an empirical approach to value selection given that certain views (e.g. those concerning social status) are widely endorsed but hard to justify from an ethical vantage point.

Concrete Problems

In practice, AI systems that have been deployed in the world or in training environments have also encountered a number of difficulties that bear upon alignment. At their root, sits a number of attributes that we have already mentioned: autonomy, intelligence, and powerful optimization-based learning. In particular, there is concern that the elements may combine in ways that lead the goal of an AI system, as established by its reward function, to diverge from its human operator's true goal or intention (Christiano, 2018; Leike 2018). As ML systems become more powerful, there are at least four specific challenges that AI researchers need to overcome (Amodei et al 2016).

The first challenge is *reward hacking* or *reward corruption*. This problem arises when an artificial agent manages to maximise the numerical reward it receives by finding unanticipated shortcuts or corrupting the feedback system (Ring and Orseau 2011). One famous example of this occurred when an RL agent was trained to play the computer game CoastRunners. In this case, an agent that had been trained to maximise its score looped around and around in circles *ad infinitum*, crashing into obstacles, collecting points, and achieving a high score – all without finishing the race, which is what it was really meant to do (Clark and Amodei, 2016). To address this challenge, researchers have considered a number of options. Everitt et al (2017), propose two strategies for promoting aligned behavior: engineers can provide the agent with richer data so that it avoids mistakes arising from 'systemic sensory error', and they can blunt some of the force of reward hacking when it stems from strong forms of optimization by focusing on a random sample of top-tier outcomes instead of a single specific objective. Other important approaches include building agents that always entertain a degree of uncertainty about their true reward, something that creates an incentive to consult human operators and ensure that their current policy is still on-track (Hadfield-Menell, 2016; Russell, 2019).

Second, even if the agent aims for the right goal or outcome, it may simply take the most efficient path to that goal and not factor in *negative side effects*. To avoid this outcome, researchers have focused on ways to ensure artificial agents are not significantly disruptive, when compared to a counterfactual baseline, and also that they demonstrate 'conservatism' in this regard by minimizing irreversible changes to the environment (Krakovna, 2019; Turner, 2019). More advanced approaches aim to ensure that agents really understand the meaning of the instructions they are given – including assumptions about outcomes to avoid – by drawing upon contract theory (Hadfield-Menell, 2018). The key here is to provide agents with access to 'substantial amounts of external structure' that supplement and fill in any gaps that are left by explicit statements of the goal.

Third, in order to learn successful policies, agents need to engage in a process of exploration where they try different things and receive feedback from the environment. As a consequence, we need to make sure that agents *explore the world in a safe way* and that they do not make costly mistakes while doing so. Often, it makes sense to train agents in simulation and allow them to make mistakes in a contained setting. However, an important element of alignment research centres upon testing agents and ensuring that they are able to perform well in the wild.

Finally, there is the challenge of how to assess and *evaluate complex agent behavior*. We have already noted that algorithms are often opaque, leading some commentators to use the metaphor of a 'black box' (Pasquale, 2015). These qualities are particularly challenging in social and legal contexts where those affected are entitled to an explanation of why decisions were taken. However, a related problem also occurs at the micro-level when it comes to training and evaluating artificial agents. Usually, this process requires a lot of feedback that is costly to give, in terms of time and attention. As a consequence, designers may use more simple proxies to evaluate agent behavior. For example, we might check only for visible dirt when evaluating the performance of a cleaning robot, rather than checking under surfaces or doing a full evaluation. However, these proxies then make it more likely that the agent will drift off track or engage in faulty behavior. This challenge becomes more complicated still when we think about training and interacting with very advanced artificial agents.

Highly Advanced AI

Looking forward, a number of theorists have suggested that AI systems are likely to become more powerful in the future. Stuart Russell describes the 'ultimate goal' of AI research as the discovery of a general-purpose 'method that is applicable across all problem types and works effectively for large and difficult instances while making very few assumptions' (2019, 46). Among experts working in this area, AI that matches or exceeds human-level intelligence across different domains is often referred to as 'artificial general intelligence' (AGI). This notion is closely related to and sometimes equated with the idea of 'superintelligence' which Nick Bostrom defines as 'any intellect that greatly exceeds the cognitive performance of humans in virtually all domains of interest' (2014, 26).

Importantly, while the creation of superintelligence could, presumably, unlock great benefits for humanity, Bostrom has argued that it also poses an existential risk in cases where its objectives are not aligned with our own (Bostrom 2014). Moreover, catastrophic outcomes need not be the result of malice on the part of the agent or its designers: rather, they could be a by-product of other tendencies or inclinations that such an agent might have. At the cornerstone of Bostrom's argument sits the orthogonality thesis which holds that any level of intelligence is compatible with any goal. If it is correct, then we should not expect there to be any correlation between how intelligent a machine is and how closely it is aligned with our values. More specifically, the capacity for instrumental rationality – that AI systems exemplify to a high degree – does not equate to alignment with certain substantive goals or outcomes. In this regard, Bostrom is at odds with Derek Parfit (2011) and Peter Singer (2011), who hope that substantive moral insight might result from the capacity for instrumental reason, and has more in common with David Hume who held that instrumental rationality is compatible with any final end (1739-40/2000).

Second, both Bostrom and Russell defend a version of the *instrumental convergence thesis*, which predicts that AGI would display instrumental goals of self-improvement, self-preservation, and resource acquisition in pursuit of its goals, even if this is to the disadvantage of human beings. As Russell points out, 'any entity that has a definite objective will automatically act as if it also has instrumental goals' (2019, 141-42). One such instrumental goal is for the machine to stay switched on so it can fulfill other objectives [the off-switch problem']; others might be acquiring money or pursuing 'resource objectives' such as computing power, algorithms, and knowledge since these are 'useful for achieving any overarching objective.' The problem, Russell suggests, is that 'the acquisition process will continue without limit', thereby necessarily creating a conflict with human interests and needs (ibid.). As with more prosaic examples of reward-hacking, Russell's solution is to ensure that AGI is *uncertain* about its true objectives so that it exhibits 'a kind of humility', exemplified in behavior such as deferring to humans and allowing itself to be switched off.

Third, even if the previous two challenges can be surmounted there is still a challenge around the *intelligibility and supervision* of AGI. The central question here is how to provide scalable advice and direction to an entity whose capabilities, knowledge, and action space are, in certain respects, beyond our comprehension. Most approaches to this task aim to break down the normative evaluation of agent behavior into smaller tasks, so that humans can then add up their evaluations and arrive at an overall judgement – even when artificial agents are pursuing complex goals on a vast scale. In this context, *reward modelling* is a set of techniques for supplementing RL with a learned reward function, trained with human oversight and monitoring (Leike et al 2018). In 'recursive reward modeling', the artificial agent is specifically incentivized to help its human instructor better define goals for the AI to pursue, enabling more effective evaluation of the agent's behavior even as it scales up. Recursive reward modeling is an example of *iterated amplification*, an approach that trains the AI progressively by breaking down a task into simpler sub-tasks (Christiano et al 2018). Lastly, *safety via debate* is a modified version of iterated amplification which involves training systems that debate with each other, competing to provide true answers to human operators (Irving et al 2018). Given a question or proposed action, two AI agents take turns making short statements, enabling a human judge to decide which gave the most useful information.

5. The Fundamental Relevance of Value

We have now covered a significant amount of ground, from the technology-value nexus, to the properties of AI systems, to research into agents whose capabilities might one day significantly exceed our own. With this architecture in place, are there any new insights that we can draw upon for the purpose of AI alignment? In particular, can long standing discussion of the relationship between technology, values and society, help illuminate this debate? We believe it can, both in terms of alignment's ultimate goals and in terms of how these outcomes should be brought about.

Given the understanding of intelligence used in this domain, artificial agents will necessarily pursue *some* goal or objective. This then raises normative questions about *what* kind of goal or objective AI systems should be designed to pursue. Within the AI research community there are three prominent answers to the question 'alignment with what?' The first approach focuses on alignment with *instructions*, aiming to include as much safety or value-preserving information as possible in the orders that the AI system receives. Russell (2019) refers to this as the 'standard model'. However, he points out that instructions may be understood very literally by agents that lack contextual understanding, and that this could have negative consequences – with the story of King Midas serving as an illustration. This risk has led some researchers to focus instead on creating agents that behave in accordance with the user's true *intentions*. Indeed, this locus for alignment sits at the heart of the reward modelling approach discussed earlier, and also lends weight to the notion of using contract theory to ensure that AI understands the implied meaning of terms (Hadfield-Menell, 2018). A third approach, which is endorsed by Russell among others, aims to align artificial agents with *human preferences*, something that might be achieved using IRL.

However, viewed from the standpoint of the technology-value nexus, there is clear potential for a gap to open up between each of these loci for alignment and the values or states of affairs that a technology ultimately helps us realise. After all, instructions, intentions and preferences, can all be misinformed, irrational, or unethical, in which case their promotion by AI would lead to bad states of affairs. The aspiration to create agents that are aligned with values – which is to say the full spectrum of things that we should promote or engage with – opens up a number of questions about how this could be done.

Additionally, alignment research has tended to focus on aligning AI with the instructions, intentions, or preferences, of a single human operator. In part because of the sizable technical challenges we have considered, discussion of alignment has tended to centre upon scenarios that are 'one-to-one' rather than 'one-to-many'. As a consequence, less attention has been paid to the question of how these various elements can be integrated or made to work on a society-wide or even global basis. To the extent that these questions have arisen for more general AI systems, there have been two main kinds of proposal: one of which focuses on social choice theory (Prasad 2018; Baum 2017) and the other of which focuses on the possibility of ideal convergence between different perspectives and opinions (Yudkowsky 2004). The first approach has been studied extensively in the domain of welfare economics and voting systems. It looks at how to aggregate information from individual people into collective judgements. In the context of value alignment, Stuart

Armstrong (2019) has argued that these approaches could be used to systematically synthesize different types of human preference (including basic preferences about the world and meta-preferences) into a utility function that would then guide agent behavior. The second approach has been discussed by Eliezer Yudkowsky (2004) who suggests that AI could be designed to align with our 'coherent extrapolated volition'. This goal represents an idealized version of what we would want 'if we knew more, thought faster, were more the people we wished we were, and had grown up farther together' (2004).

However, this focus on 'one-to-one' versions of alignment and emphasis on preference aggregation, potentially elide important aspects of the alignment question. To start with, a fuller appreciation of the social consequences of technology, and the way in which it shapes our own choice architecture, points toward a need for richer and more democratic forms of AI alignment. To achieve what we have termed 'social value alignment', of the kind often advocated for by the fairness, accountability and transparency community, AI systems ultimately need to embody principles that are widely endorsed by those who have their lives affected – and sometimes powerfully so – by these technologies.

Alignment with instructions and intentions, or the capability to understand human preferences, may still be an important stepping stone towards this goal. However, the richer ideal of social value alignment itself presses us to engage with a new set of challenges, including the challenge of moral uncertainty (i.e. the fact that we are often unsure what action or theory is morally right) and the challenge of moral pluralism (i.e. the fact that people ascribe to a variety of different reasonable views and perspectives). Taken together, these elements mean that we are unlikely to persuade everyone about the truth of a single moral theory using evidence and reason alone (Rawls 2003). As a consequence, if we are to avoid a situation in which some people simply impose their views on others, then AI alignment necessarily has a social and political dimension: we need to collectively come up with principles or values that we agree are appropriate for this purpose (Gabriel, 2020).

In practice, many public and private bodies have sought to come up with principles for AI systems. These ethics guidelines tend to converge around the values of 'transparency, justice and fairness, non-maleficence, responsibility and privacy' (Jobin, 2019). However, they have also led to a well-founded concern that the voices included in these processes are not truly representative of affected parties. Some researchers have argued, for instance, that high-level AI values statements tend to promote a 'limited, technologically deterministic, expert-driven view', setting the terms of debate in a way that make 'some conversations about ethical design possible while forestalling alternative visions' (Greene at el 2019, 2122). Indeed, the prominent AI researcher Shakir Mohamed has expressed concern that AI research is currently 'localised' and '[w]ithin restricted geographies and people' (2018). He writes that 'We [AI researchers] rely on inherited thinking and sets of unquestioned values; we reinforce selective histories; we fail to consider our technology's impacts and the possibility of alternative paths; we consider our work to be universally beneficial, needed and welcomed' (ibid).

Moving forward, these considerations point toward the need for a deepening conversation about the nature of AI alignment including what it means for agents to be socially aligned in different contexts. If the technology comes to have increasingly global reach then one aspiration might be to build alignment around principles that are subject to a 'global overlapping consensus' (Gabriel, 2020). These principles would foreground commonalities between diverse systems of thought and could, in principle, be endorsed by the wide range of people who these technologies affect.[8] However, consensus of this kind needs to be the result of deep and inclusive discussion if it is to have real value (Shahriari, 2017). This is for two reasons. First, those affected have a right to contribute to discussions about technologies that have a profound effect on them. The legitimacy of the resulting principles depends upon antecedent recognition of the right to speak about and influence these matters. Second, regarding the epistemic value of the principles themselves, it is important to remember that no individual has a complete monopoly on the truth. In this regard, it is far more likely that J.S. Mill was correct when he suggested that everyone only has access to a part of it (Mill, 1859/2006, 53). By creating an open and properly inclusive discourse around AI ethics we create space for new considerations to come to light, something that should lead to a richer, more complete set of guidelines and principles over the long run.

6. Conclusion

This chapter has sought to situate questions around AI alignment within wider discussion of the relationship between technology and value. We have suggested that the connection between technology and value is best understood through the impact technological artefacts have on our ability and inclination to access various states of affairs. We also argued that this relationship is potentially more complex and salient for AI than it is for simple technological artefacts, given that new agents or models embody greater degrees of freedom and can be loaded with thicker values than was true of objects in the past. Indeed, when it comes to the evaluation of AI systems, we look not only for guarantees that AI artefacts were *designed* in ways that demonstrate care or compassion, but also for AI systems to evidence these qualities themselves. This shift in the locus of evaluation reflects the fact that AI systems are often best understood from the 'intentional stance', a perspective that allows for the possibility they have qualities of the relevant kind.

Looking forward, this combination of agency and intelligence gives rise to challenges that the technical AI community seeks to address. These matters are rightfully the focus of serious research efforts given that the challenges could potentially scale to more powerful systems. At the same time, they should not overshadow the question of what values AI systems should ultimately be aligned with. Ultimately, we need to reach beyond alignment with a single human operator and think about what it means for AI technology to be *socially value aligned*. These questions are already being thought about and addressed by the fairness, accountability and transparency community, creating a significant opportunity for feedback and mutual learning. More generally, these observations highlight the importance of interdisciplinary approaches for value alignment efforts. As a relatively new area of moral and technical inquiry, there is an opportunity to harness different branches of knowledge as part of a broad and inclusive research agenda. It points also to the need for technologists and policy-makers to stay attuned to their social context and stakeholders, even as the AI systems they build become more capable. Considerations of fair process and epistemic virtue point toward the need for a properly inclusive discussion around the ethics of AI alignment.

---

[8] The challenge of ensuring different voices are properly heard has parallels in information ethics, which draws a distinction between a 'mono-cultural view of ethics' that claims exclusive validity, and a 'transcultural ethics' arising from intercultural dialogue (Capurro 2007). It also resonates with concerns raised by feminist epistemologists working on the philosophy of science, including the downplayinng of certain cognitive styles and modes of knowledge, and the production of technologies that reinforce gender and other social hierarchies (Anderson 2015).

7. <u>Bibliography</u>